\documentstyle[12pt,epsf,rotate]{article}
\input FEYNMAN
\newcommand{\pr}{\paragraph{}}
\newcommand{\nn}{\nonumber}
\newcommand{\be}{\begin{equation}}
\newcommand{\ee}{\end{equation}}
\newcommand{\bea}{\begin{eqnarray}}
\newcommand{\eea}{\end{eqnarray}}

\begin{document}
\newcommand{\nd}[1]{/\hspace{-0.5em} #1}
\begin{titlepage}

\begin{flushright}
OUTP-97-68P \\
UA-NPPS 8/97 \\
cond-mat/9711257 \\
\end{flushright}

\begin{centering}
\vspace{.05in}
{\Large {\bf N=1 Supersymmetric Spin-Charge Separation 
in
effective gauge theories of 
planar magnetic superconductors \\}}
 
\vspace{.2in}
{\bf G.A. Diamandis}$^{a}$, {\bf B.C. Georgalas}$^{a}$ and 
{\bf N.E. Mavromatos}$^{*}$ \\
\vspace{.2in}
University of Oxford, Department of (Theoretical) Physics, 
1 Keble Road OX1 3NP, Oxford, U.K. \\

\vspace{.2in}
{\bf Abstract} \\
\vspace{.05in}
\end{centering}
{\small 
We present a $N=1$ Supersymmetric extension of a spin-charge separated
effective $SU(2)\times U_S(1)$ `particle-hole' 
gauge theory of excitations about the nodes of the gap of 
a $d$-wave  
planar magnetic superconductor. The supersymmetry is achieved 
without introducing extra
degrees of freedom, as compared to the non-supersymmetric models.
The only exception, the introduction of gaugino fieds, 
finds a natural physical 
interpretation as describing 
interlayer coupling in the statistical model.  
The low-energy continuum theory is 
described by a relativistic 
(2+1)-dimensional supersymmetric $CP^1$ $\sigma$-model with 
Gross-Neveu-Thirring-type four-fermion interactions.
We emphasize 
the crucial  r\^ole of the $CP^1$ constraint in 
inducing a non-trivial dynamical mass generation for 
fermions (and thus superconductivity), in a way 
compatible with manifest $N=1$ supersymmetry. 
We also give 
a preliminary discussion of non-perturbative effects.
We argue that supersymmetry 
suppresses the dangerous for superconductivity 
instanton contributions 
to the mass of 
the perturbatively massless gauge boson
of the unbroken $U(1) $ subgroup of $SU(2)$. 
Finally, we point out the possibility of applying 
these ideas to effective gauge models of 
spin-charge separation in  
one-space dimensional 
superconducting chains of holons, which, for example, have recently been 
claimed to 
be important in  
the stripe phase 
of underdoped cuprates.}

\vspace{0.6in}

\begin{flushleft} 
November 1997 \\
$^{a}$ On leave from University of Athens, 
Physics Department,
Elementary Particle and Nuclear Physics Section, 
Athens GR-157 71, 
Greece. \\
$^{(*)}$ P.P.A.R.C. Advanced Fellow. 
\end{flushleft} 

\end{titlepage}

\section{Introduction} 

In ref.~\cite{fm} 
it was argued that the doped 
large-$U$ Hubbard (antiferromagnetic) models possess
a {\it hidden} local {\it non-Abelian} $SU(2)\times U_S(1)$ 
phase symmetry related to spin interactions.
This symmetry 
was discovered using an appropriate `particle-hole symmetric
formalism' for the electron operators~\cite{zou}, 
and employing a generalized 
{\it slave-fermion} ansatz for {\it spin-charge 
separation}, 
which allows intersublattice 
hopping for holons, viewed as fermions. 
The spin-charge separation 
may be physically interpreted as implying an effective 
`substructure' of the electrons due to the 
many body interactions in the medium. 
This sort of idea, originating from Anderson's 
RVB theory of spinons and holons~\cite{Anderson},
was also pursued recently by 
Laughlin, although from a (formally at least)
different perspective~\cite{quark}.

In ref. \cite{fm} we have argued in favour of 
the opening of a fermion gap at the nodes of a $d$-wave gap
of a superconducting antiferromagnet. 
Linearization of the fermion spectrum about such nodes 
leads to a relativistic Dirac spectrum for holons, with the r\^ole of the 
limiting velocity being played by the fermi velocity~\cite{dor,fm}.
Such systems might be 
of
relevance to the physics of high-temperature
superconductors, since recently it is believed that 
high-temperature superconductivity in cuprates 
is highly anisotropic and the gap symmetry is of $d$-wave type~\cite{dwave},
with the gap vanishing along lines of {\it nodes} 
on the Fermi surface~\footnote{There is also recent experimental 
evidence on the possibility of the opening of a gap at
such nodes, triggered by either magnetic fields~\cite{ong} 
or by magnetic impurities~\cite{recent}, and although
such phenomena
might admit alternative (more conventional) 
explanations~\cite{laughlinT,balatsky}, however
the r\^ole of spin-charge separation in this context 
still remains a challenging project.}. 

The key suggestion in ref. \cite{fm}, which lead to 
the non-abelian gauge symmetry structure for the doped antiferromagnet,
with the constraint 
of {\it not more than one electron per lattice site}, 
was the {\it slave-fermion} spin-charge separation ansatz 
for physical electron operators~\cite{fm}: 
\be
\chi _{\alpha\beta,i} 
\equiv \left(
\begin{array}{cc}
c_1 \qquad c_2 \\
c_2^\dagger \qquad -c_1^\dagger \end{array}
\right)_i \equiv {\widehat \psi} _{\alpha\gamma,i}{\widehat z}_{\gamma\beta,i} =
\left(\begin{array}{cc}
\psi_1 \qquad \psi_2 \\
-\psi_2^\dagger \qquad \psi_1^\dagger \end{array}
\right)_i~\left(\begin{array}{cc} z_1 \qquad -{\overline z}_2 \\
z_2 \qquad {\overline z}_1 \end{array} \right)_i 
\label{ansatz2}
\ee
where $i$ is a {\it lattice site} 
index,
$c_\alpha$, $\alpha=1,2$ are electron annihilation
operators, the Grassmann variables $\psi_i$, $i=1,2$ 
play the r\^ole of holon excitations, while the bosonic
fields $z_i, i=1,2,$ represent magnon (bosonized spinon) 
excitations~\cite{Anderson}.
The ansatz (\ref{ansatz2}) 
has spin-electric-charge separation, since only the 
fields $\psi_i$ carry {\it electric} charge.

As argued in ref. \cite{fm} 
the ansatz is characterised by   
the following {\it hidden local
phase} (gauge) symmetry structure: 
\be
  G=SU(2)\times U_S(1) \times U_E(1) 
\label{group}
\ee
The 
gauge 
SU(2) symmetry 
pertains to the spin degrees of freedom.
The 
local $U_S (1)$ `statistical' phase symmetry  
allows fractional statistics of the spin and charge 
excitations. This is an exclusive feature
of the three dimensional geometry, and is similar in spirit
to the bosonization technique of the spin-charge 
separation ansatz of ref. \cite{marchetti}.
Finally the $U_E(1)$ symmetry is due to the electric
charge of the holons. 

It is the purpose of this work to 
discuss the possibility of a 
{\it hidden supersymmetry} in the ansatz (\ref{ansatz2}). 
Note that 
supersymmetric extensions of $J=\pm 2t$ models for doped antiferromagnets,
in one and two spatial dimensions, 
have already appeared in the existing literature~\cite{susytj}, 
even in the context of spin-charge separated anyon 
models~\cite{susytj2}.
However, as far as we are aware, 
such supersymmetries have not been associated so far with 
any specific dynamical properties of the antiferromagnet. 
In contrast,  
in our 
approach here, based on the non-trivial ansatz (\ref{ansatz2}),
the 
supersymmetry constitutes a non-trivial {\it dynamical} property of 
the  spin-charge separated vacuum
for {\it holons} and {\it spinons}, by viewing them 
as {\it supersymmetric partners}. 
Due to the rich group structure (\ref{group}),
many possibilities arise in the study of the phase diagrams
of these theories, in the context of the modern perspective  
advocated in the work of Seiberg and Witten~\cite{seiberg,intrilligator}. 
In particular, duality symmetries in
the infrared region of the supersymmetric 
model, connecting various theories
with the same non-trivial infrared fixed-point~\cite{intrilligator},
may prove very useful in a renormalization-group study of the 
dynamics of the gauge fields in both, the superconducting and 
the normal phases of the model, in the spirit of ref. \cite{aitch}.  
The important issue is that 
the introduction of $N=1$ supesymmetry, {\it hidden } 
in the spin-charge separation ansatz (\ref{ansatz2}), 
does not require the introduction of unphysical degrees of freedom. 
As we shall see, the only extra degrees of freedom, 
as compared to the non-supersymmetric case~\cite{fm}, 
are the gauginos 
of the local hidden gauge symmetry, which, however,  
admit the natural 
interpretation of describing {\it interlayer } hopping of 
spin {\it and} charge degrees of freedom (hopping of `real' electrons).  

We should stress that, within a condensed-matter context, 
the supersymmetry refers to the relativistic field theory at the nodes of 
a $d$-wave superconducting gap~\footnote{Galilean supersymmetry, 
as symmetry of the spectrum between bosonic and fermionic degrees of freedom,
may also occur away from the nodes. This is left for future work.}. 
In this sense, the supersymmetric 
dynamics of the spinons and holons would require
equality of the spin gap with the fermion (superconducting) gap at 
such nodes. 
At a microscopic level, this would imply some particular relation 
among the microscopic parameters of the model, such as hopping matrix elements 
and Heisenberg interactions. This calls for comparison with the $J=\pm 2t$ 
special point, where the graded (supersymmetric) 
algebra in the spectrum of the doped antiferromagnets
appears~\cite{susytj,susytj2}. However, as we shall see, the situation 
in our case is more complicated, since there are more parameters
entering the dynamical scenario of the gauge theory 
based on the spin-charge separation ansatz (\ref{ansatz2}).  

\section{Review of the (continuum) model and its superconducting properties}

Before embarking to a description of the supersymmetric extension 
we consider it as useful to review first 
the properties of the statistical model 
of ref. \cite{fm}, some of which will be crucial for 
the supersymmetric extension. 
The pertinent long-wavelength gauge model,  
describing the low-energy 
dynamics
of the large-U Hubbard 
antiferromagnet in the spin-charge separation phase (\ref{ansatz2}),
can be cast 
in a 
conventional relativistic lattice gauge-theory,   
provided 
one changes representation 
of the $SU(2)$ group, and, instead of working with $2 \times 2$ 
matrices, one uses a representation 
in which the fermionic matrices ${\widehat \psi}_{\alpha\beta}$ 
are represented as two-component (Dirac) spinors in `colour' space:
\be
{\tilde \Psi}_{1,i}^\dagger =\left(\psi_1~~-\psi_2^\dagger\right)_i,~~~~
{\tilde \Psi} _{2,i}^\dagger=\left(\psi_2~~\psi_1^\dagger
\right)_i,~~~~~i={\rm Lattice~site} 
\label{twospinors}
\ee
By assuming  
a background $U_S(1)$ 
field of flux $\pi$ per lattice plaquette~\cite{dor},
and considering quantum fluctuations around this background
for the $U_S(1)$ gauge field, 
one can obtain the conventional lattice Dirac action 
for the fermion excitations 
about a node in the fermi surface~\cite{dor,dorstat,fm}. 

In the above context, a strongly coupled
$U_S(1)$ group can dynamically generate a mass gap 
in the holon spectrum~\cite{app,dor,kocic,koutsoumbas,maris}, 
which breaks the $SU(2)$ local symmetry
down to its Abelian subgroup generated by 
the $\sigma_3$ Pauli matrix~\cite{fm,farak}.  
{}From the view point of the statistical model of ref. \cite{fm},   
the breaking of the $SU(2)$ symmetry 
may be interpreted as  
restricting the holon hopping effectively to 
a single sublattice, since 
the intersublattice hopping is suppressed 
by the mass of the gauge bosons.  

The (naive) continuum limit of the low-energy theory 
about such nodes on the fermi surface of the 
planar antiferromagnet, then, 
is described by a $CP^1$ model coupled to Dirac fermions~\cite{dor,fm}:
\be 
{\cal L}_2 = g_1^2|(\partial _\mu 
- (g_2/g_1)B_\mu ^a \sigma ^a - a_\mu  )z|^2  
+i{\overline \Psi}D_\mu\gamma_\mu\Psi
\label{su2action}
\ee
where now $D_\mu = \partial_\mu -ia_\mu-i(g_2/g_1)\sigma^aB_{a,\mu}
-\frac{e}{c}A_\mu$, $g_i^2$, $i=1,2$ have dimensions of mass, 
$B_\mu^a$ is the gauge potential of the local (`spin') $SU(2)$ group,
generated (in two-component notation for fermions ) by the Pauli 
matrices $\sigma ^a$,
$a_\mu$ the $U_S(1)$ (`fractional statistics') field, and 
$A_\mu$ is an external electromagnetic potential, which 
will be ignored in the subsequent discussion. 
In terms of the microscopic model, $g_1^2 \sim J \delta $, where 
$J$ is the Heisenberg exchange energy, and $\delta $ is the doping 
concentration. 
An important ingredient in the above formalism is the no-double 
occupancy 
constraint, which in terms of the $z$ and $\Psi _\alpha$, $\alpha=1,2$, fields,
with $\alpha$ a `colour' index, can be written as: 
\be
\sum_{\alpha=1}^{2} [ {\overline z}^\alpha z _\alpha + \beta 
{\overline \Psi}^{\alpha} \sigma _3 \Psi _\alpha ] = 1 
\label{constr}
\ee
where $\sigma _3 $ acts in spinor space, and  
the fermions $\Psi$ are viewed as 
{\it two-component} spinors, related to the spinors 
${\tilde \Psi }$ (\ref{twospinors}) by 
appropriate rescalings so as to ensure the canonical 
kinetic (Dirac) term~\footnote{In the model 
of ref. \cite{fm}, due to the Dirac nature of the resulting spinors, 
$\Psi ^\dagger$ and $ \Psi $ are viewed as independent variables
in a path integral, which implies that one can redefine 
$\Psi ^\dagger \rightarrow {\overline \Psi }$.}. 
This results in the presence of the 
constant $\beta $ (with dimensions of [mass]$^{-2}$) 
in the constraint (\ref{constr})~\cite{dorstat}.  
In the context of the microscopic model,
these constants  are expressed  
in terms of the hopping and Heisenberg exchange 
energies~\cite{dorstat,fm}, and one has that $|\beta | << 1$. 
It can be shown~\cite{fm} that the constraint (\ref{constr}) 
is essential 
in ensuring the consistency of the ansatz (\ref{ansatz2}) with the 
canonical commutation relations of the electron operators. 

The presence of the $\Psi ^\dagger \Psi $ (non-relativistic) 
fermion number term in the constraint (\ref{constr}) 
appears at first sight to complicate things, since  
the conventional $CP^{1}$ constraint 
$|z|^2 = 1$ is no longer valid. 
However, these extra terms can be rendered inocuous for the 
dynamics of the effective theory. 
Indeed, 
by integrating out the 
(non-propagating ) gauge fields in (\ref{su2action}) we obtain~\cite{ruiz}:  
\bea
&~& {\cal L}_{B} = g_1^2 \partial ^\mu {\overline z}^\alpha 
\partial _\mu z_\alpha + i{\overline \Psi}^\alpha 
\nd{\partial} \Psi _\alpha 
+ \nn \\
&~& \frac{g_1^2}{2}
[1-\beta {\overline \Psi}^{\alpha}\sigma _3 \Psi_\alpha ]^{-1} 
{\rm Tr} \left({\overline z}^\alpha \partial _\mu z_\alpha - 
z_\alpha \partial _\mu {\overline z}^\alpha - i g_1^{-2}{\overline \Psi}^\alpha 
\gamma _\mu (1 + \sigma^a)\Psi _\alpha \right)^2
+ \nn \\
&~&6{\rm ln}[1-\beta {\overline \Psi}^{\alpha} \sigma _3 \Psi _\alpha ] 
\label{lagrangianb}
\eea
where the last term is absent in the usual $CP^1$ models. 
Expanding this term in powers of the (small) parameter $\beta << 1$, 
one obtains: 
\be
6{\rm ln}[1-\beta {\overline \Psi}^{\alpha} \sigma _3 \Psi _\alpha ] \simeq
-6\beta {\overline \Psi} ^{\alpha} \sigma _3 \Psi _\alpha 
+ 3\beta ^2 ({\overline \Psi}^{\alpha}\sigma _3 \Psi _\alpha )^2 + \dots
\label{expansion}
\ee
where the $\dots $ indicate six- and higher order -fermion contact terms, 
not renormalizable, even in large-N limits, which constitute irrelevant 
operators, in a renormalization-group sense, not affecting the 
low-energy (infrared) structure of the effective theory, we are 
interested in. 

Applying a Hartree-Fock linearization to the four-fermion interactions, 
one obtains 
terms of the form: 
\be 
 3\beta ^2 <{\overline \Psi}^{\beta} \sigma _3 \Psi _\beta > 
{\overline \Psi}^{\alpha} \sigma _3 \Psi _\alpha 
 \label{condensate}
\ee
Collecting the $\Psi ^\dagger \Psi $ terms together,  
one then obtains   
a fermion-density term in the effective lagrangian of the form: 
\be
  L_\mu = (-6\beta + 3\beta ^2 <{\overline \Psi} ^{\alpha} \sigma _3
\Psi _{\alpha} > )
{\overline \Psi} ^{\beta } \sigma _3 \Psi _{\beta} 
\label{chmical}
\ee
Upon inserting the constraint (\ref{constr}) via a Lagrange multiplier field 
$\lambda (x)$ in the path integral, one may expand~\cite{Polyakov} 
about the vacuum defined by $< \lambda (x) > \propto m_Z^2 \ne 0$, where $m_Z$ 
is a spinon gap (magnon mass), in appropriate units. Then, 
we can tune the parameter of our system 
so as to define a fully relativistic field theory about the nodes of a 
$d$-wave gap~\cite{fm}, such that, when $\beta \ne 0$, the following 
is satisfied: 
\be 
   <\lambda (x) >\beta  
-6\beta + 3\beta ^2 < {\overline \Psi}^\alpha \sigma _3 
\Psi _\alpha > = 0 
\label{absence}
\ee
Note that the non-zero dynamical condensate of the (non-relativistic) 
operator $<{\overline \Psi}^\alpha \sigma _3 \Psi _\alpha> $, 
obtained above, is compatible 
with a dynamical opening of a fermion mass gap in the resulting relativistic 
field theory.  

In this way, the fermion terms in the constraint 
(\ref{constr}) decouple, and 
the effective theory of the excitations 
at the nodes of the $d$-wave superconducting gap 
can be described, up to terms that are renormalization-group
irrelevant 
operators
in the infrared, 
by the effective theory (\ref{lagrangianb}) with 
Thirring four-fermion interactions, and a standard 
$CP^1$ constraint: 
\be
\sum _{\alpha=1}^{2} |z^\alpha|^2 = 1 
\label{cpconstr}
\ee
The latter implies that 
the magnon fields, in their massive (spin gap) phase, 
can be integrated out in a standard
fashion in the path integral~\cite{Polyakov}, 
to yield an alternative low-energy theory, that of
a dynamical $SU(2) \times U_S(1)$ gauge group, with Maxwell kinetic terms
for the gauge fields, which are the dominant terms in a derivative
expansion. 

Superconductivity in this model occurs~\cite{fm} 
as a result of dynamical generation of a parity-conserving fermion 
mass in the strong-coupling regime of 
the $U_S(1)$ gauge field~\cite{app,dor,fm}, upon coupling 
the system to external electromagnetic potentials. 
This dynamical generation phenomenon occurs in the infrared region
of the effective theory obtained after $z$-magnon integration. 
In such a theory, upon the 
opening of a mass gap in the fermion (holon) spectrum, 
the
Feynman matrix element: 
${\cal S}^a = <B^a_\mu|J_\nu|0>, a=1,2,3$, with $J_\mu ={\overline
\Psi}\gamma _\mu \Psi $ the fermion-number current, is non-trivial.
Due  
to the colour-group structure, only the massless $B^3_\mu $ 
gauge boson of the $SU(2)$ group, corresponding to the $\sigma _3$
generator in two-component notation, contributes to the 
matrix element. 
The non-trivial result for ${\cal S}^3$
arises from an {\it anomalous one-loop graph}, depicted in 
figure 1, and it
is given by~\cite{RK,dor}:
\be
    {\cal S}^3 = <B^3_\mu|J_\nu|0>=({\rm sgn}{m_f})\epsilon_{\mu\nu\rho}
\frac{p_\rho}{\sqrt{p_0}} 
\label{matrix2}
\ee
where $m_f$ is the parity-conserving fermion mass, generated dynamically 
by the $U_S(1)$ group. As with the 
other Adler-Bell-Jackiw anomalous graphs in gauge theories, 
the one-loop result (\ref{matrix2}) 
is {\it exact} and receives no contributions from higher loops~\cite{RK}.

\begin{centering}
\begin{figure}[htb]
\vspace{2cm}
%
\bigphotons
\begin{picture}(30000,5000)(0,0)
\put(20000,0){\circle{100000}}
\drawline\photon[\E\REG](22000,0)[4]
\put(18000,0){\circle*{1000}}
\end{picture}
%
\vspace{1cm}
\caption{{\it Anomalous one-loop Feynman matrix element,
leading to a Kosterlitz-Thouless-like breaking of the 
electromagnetic $U_{E}(1)$ symmetry, and thus 
superconductivity, once a fermion 
mass gap opens up. The wavy line represents the $SU(2)$ 
gauge boson $B_\mu^3$,
which remains massless, while the blob denotes an insertion 
of the fermion-number
current  $J_\mu={\overline \Psi}\gamma_\mu \Psi$.
Continuous lines represent fermions.}}
\label{fig2}
\end{figure}
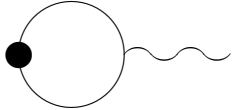
\end{centering}
\pr
This unconventional 
symmetry breaking (\ref{matrix2}), 
does {\it not have a local order parameter}~\cite{RK,dor},
since the latter is inflicted by strong phase fluctuations, 
thereby resembling  the
Kosterlitz-Thouless
mode of symmetry breaking\footnote{This may be important from a 
condensed-matter viewpoint, since 
the absense of a local order parameter 
implies that
the opening of a fermion mass 
gap at the {\it nodes} of the original $d$-wave 
superconducting gap of the cuprate
does not affect the $d$-wave
nature of the state.}. The {\it massless} gauge boson 
$B_\mu^3$ of the 
unbroken $\sigma_3-U(1)$ subgroup of $SU(2)$ is responsible for the 
appearance of a {\it massless pole} in the electric current-current 
correlator~\cite{dor}, which is the characteristic feature 
of any {\it superconducting theory}. As discussed in ref. \cite{dor},
all the standard properties of a superconductor, such as 
the Meissner
effect, infinite conductivity, flux quantization, London action etc. are 
recovered in such a case. 
The field $B^3_\mu$, or rather its {\it dual} $\phi$ defined by
$\partial _\mu \phi \equiv \epsilon_{\mu\nu\rho}\partial_\nu B^3_\rho$,
can be identified with the Goldstone
boson of the broken $U_{em}(1)$ (electromagnetic) symmetry~\cite{dor}. 
We shall come back to the exactness of this result, upon including 
non-perturbative effects (instantons), in the context of our supersymmetric 
model, later on. 

\section{N=1 Supersymmetric Gauge Theory of Spin-Charge Separation} 

We are now ready to 
discuss 
the possibility of the 
emergence of a $N=1$ space-time supersymmetry in 
the ansatz (\ref{ansatz2}). 
The main idea behind such a supersymmetrization is to view 
the magnons $z$ as {\it supersymmetric partners} of the holons $\Psi$.
For simplicity, in this note we shall turn off the $SU(2)$ interactions 
in (\ref{su2action}), keeping only $U_S(1)$, which is mainly responsible 
for the chiral symmetry breaking (mass generation) phenomenon. 
The 
incorporation of the gauge 
group $SU(2) \times U_S(1)$ (\ref{group}) is straightforward. 
In this section we shall demonstrate the possibility of 
a $N=1$-supersymmetric extension  of 
the action (\ref{su2action}), and of the constraint
(\ref{constr}), in the absence of 
(non-supersymmetric) external electromagnetic potentials. 

The basic ``matter" multiplet of N=1 supersymmetry in three 
space-time dimensions,
can be written in terms of a scalar superfield as~\cite{harvey} 
\be
\Phi = \phi + \bar{\theta} \chi + \frac{1}{2} \bar{\theta} \theta F
\label{scalars}
\ee
which contains a real scalar field, $\phi$, a Majorana spinor
$\chi_\alpha$ and a real auxiliary field $F$.
We consider complex superfields
\be
Z = \frac{1}{\sqrt2}(\Phi_1 + i\Phi_2) = 
z + \bar{\theta} \psi + \frac{1}{2} \bar{\theta} \theta F
\label{complex}
\ee
which contain a complex scalar, $z=\frac{1}{\sqrt2}(\phi_1 + i\phi_2)$,
a Dirac spinor, $\Psi=\frac{1}{\sqrt2}(\psi_1 + i\psi_2)$, and a
complex auxiliary field, $F=\frac{1}{\sqrt2}(F_1 + iF_2)$.
The supersymmetry transformations read,
\bea
&~& \delta_S z = \bar{\xi} \psi \nn \\
&~& \delta_S \psi = -i \gamma^{\mu} \xi \partial_{\mu} \phi + \xi F \nn \\
&~& \delta_S F = -i\bar{\xi} \nd{\partial} \psi
\label{scaltrans}
\eea
and the supersymmetric invariant lagrangian is given by the highest
component ($\bar{\theta} \theta$) of the superfield
\be
\bar{D} Z^* DZ
\ee
where
\be
D_{\alpha} = \frac{\partial}{\partial \bar{\theta}_{\alpha}} -
i (\nd{\partial} \theta)_a
\label{covder}
\ee
is the supersymmetry covariant derivative.
\newline
The gauge field is incorporated in a real spinor superfield which,
in the Wess-Zumino gauge, takes the form
\be
V_{\alpha} = i (\nd{a} \theta)_{\alpha} + \frac{1}{2} \bar{\theta}
\theta \eta_{\alpha}
\label{vector}
\ee
where $\eta$ is the supersymmetric partner of the gauge field (gaugino).
\newline
The supersymmetric  gauge invariant lagrangian for the matter fields
which in terms of superfields is the highest component of the superfield
\be
{\overline {\cal D}} Z^* {\cal D} Z
\label{superg}
\ee
with
\be
{\cal D}_{\alpha} = D_{\alpha} - iV_{\alpha}
\label{supercov}
\ee 
In terms of component fields the lagrangian reads:
\be
L = g_1^2 [D_{\mu} \bar{z}^{\alpha}D^{\mu}z^{\alpha} +
i {\overline \Psi} \nd{D} \Psi + \bar{F}^{\alpha} F^{\alpha} 
 + 2i({\overline \eta} \Psi^{\alpha} \bar{z}^{\alpha} -
{\overline \Psi}^{\alpha} \eta z^{\alpha})]
\label{lagrangian}
\ee
where $D_\mu$ denotes the gauge covariant derivative with respect to the 
$U_S(1)$ field, and for convenience we have rescaled the 
fermion fields $\Psi $ and the auxiliary field $F$ by $g_1$, as compared 
to the non-supersymmetric case, 
in order to facilitate our superfield formalism.
Notice that (\ref{lagrangian}) contains a supersymmetric 
partner of the statistical gauge field $U_S(1)$. From the point of view of the ansatz (\ref{ansatz2}), this is the defining property of the
$N=1$ supersymmetric point of the model, in the sense that the 
gauge interaction $U_S(1)$ `doubles' its 
degrees of freedom. From the point
of view of the statistical model of ref. \cite{fm}, this doubling 
will only be reflected in the form of the effective action, after 
integrating out the $U_S(1)$ field. As explained in ref. \cite{fm}, 
this field is responsible for yielding fractional statistics 
to the holons and spinons in three dimensions, and as such should be 
integrated out in the 
effective action of the physically observable degrees of freedom. 

It is important to notice that the constraint (\ref{constr}) 
admits a $N=1$ supersymmetric formulation, in terms of the superfields 
$Z^\alpha$ (\ref{complex}):
\be
\sum _{\alpha =1}^{2}{\overline Z}^\alpha Z_\alpha = 1  
\label{sconstr}
\ee
Upon integrating out the (non-propagating)
$a_\mu$ and gaugino $\eta$ fields in a path integral for the lagrangian 
(\ref{lagrangian}), and using the constraint (\ref{sconstr}), it is immediate 
to obtain the following effective action of holons and spinons
in the spin-charge separation ansatz (\ref{ansatz2}) at 
the supersymmetric point: 
\bea
&~& {\cal L}_S = g_1^2 [\partial ^\mu {\overline z}^\alpha 
\partial _\mu z_\alpha + i{\overline \Psi}^\alpha 
\nd{\partial} \Psi _\alpha + {\overline F}^\alpha F_\alpha 
+  \frac{1}{2}
\left({\overline z}^\alpha \partial _\mu z_\alpha - 
z_\alpha \partial _\mu {\overline z}^\alpha - i {\overline \Psi}^\alpha 
\gamma _\mu \Psi _\alpha \right)^2]
+ \nn \\
&~&g_1^6 {\rm ln}\left(g_1^{-2} \sum_{\alpha}\{  
|z_\alpha|^2 {\overline \Psi}^\alpha \Psi _\alpha 
+  \sum _{\beta \ne \alpha } z_\alpha {\overline z}^\beta {\overline \Psi}^\alpha \Psi _\beta + 
\frac{1}{2} 
[z^\alpha z_\alpha {\overline \Psi}^\alpha \Psi ^*_\alpha 
+ \sum _{\beta \ne \alpha} z^\alpha z^\beta 
{\overline \Psi} _\alpha \Psi ^*_\beta + h.c.]\} \right)  
\label{slagrangian}
\eea
The auxiliary fields  $F_\alpha$ can be solved by means of the constraint
(\ref{sconstr}): 
\be
{\overline F}^\alpha F_\alpha = \frac{1}{2} \left( 
{\overline \Psi}^\alpha \Psi_{\alpha} \right)^2 
\label{ff}
\ee
The terms inside the logarithm in (\ref{slagrangian}) 
contain no bare mass terms, 
but only interaction terms 
among $z$ and $\Psi $ fields. This 
can be readily seen by the following formal expansion:   
\be 
 {\rm ln}x=2 \sum_{k=1}^{\infty} 
\frac{1}{2k -1}\left(\frac{x-1}{x+1}\right)^{2k-1}
\label{log}
\ee
which truncates due to the Grassman structures in $x$. 

Several important comments are now in order. 
First, notice that the supersymmetric extension of the effective lagrangian 
for spinons and holons contains {\it both} Gross-Neveu and Thirring 
four-fermion interactions. 
This can be seen by using the Fierz 
rearrangement formula in three space-time dimensions:
\be 
\gamma ^\mu _{ab}\gamma _{\mu,cd} = 
2 \delta _{ad}\delta _{bc} - \delta _{ab}\delta _{cd} 
\label{fierz}
\ee
upon which  
the four fermion Thirring interactions
become: 
\be 
[{\overline \Psi}_\alpha \gamma _{\mu} \Psi _\alpha ]^2= 
-3 \sum _{\alpha =1}^{2} [{\overline \Psi} _\alpha \Psi _\alpha ]^2 
- 2 [{\overline \Psi }_1 \Psi _1  {\overline \Psi }_2 \Psi _2  +
2 {\overline \Psi} _1 \Psi _2  {\overline \Psi}_2 \Psi _1 ] 
\label{thirring} 
\ee
showing that the Gross-Neveu terms in the Thirring interactions cannot cancel
the ones appearing in (\ref{slagrangian}) due to the 
supersymmetric extension.  

As a result of supersymmetry, the
couplings of the four-fermion terms are all related, and are 
of order $g_1^2$. In the context of the statistical model, such 
a restriction will imply special relations among the microscopic 
parameters, such as hopping elements, Heisenberg exchange energies, doping 
concentration etc. For instance, in the special case of ref. \cite{dorstat},
where the next-to-nearest-neighbour (NNN) hopping element 
$t'$ is assumed dominant, 
with $t \sim 0$,  one 
can show that four-fermion Gross-Neveu type terms come with generic 
coefficients of order 
$(t')^2/(J'\delta ^2)$, with $J'~(<< J)$ the NNN Heisenberg exchange energy. 
In such a situation, supersymmetry enforces the relation 
$t' \sim \sqrt{JJ'}\delta ^{3/2} $, which may be interpreted as 
implying that supersymmetric points in our formalism may be obtained
by tuning the doping concentration.  
Such restrictions may be compatible with the $t'=J'/2$ supersymmetric 
point 
of ref. \cite{susytj,susytj2}). In the case above, such an extra restriction
implies underdoped situation $\delta ^{3/2} \sim \sqrt{J'/4J} <<1$.  
In more realistic models, like the one discussed in ref. \cite{fm},
involving nearest-neighbour hopping, 
there will be  
more constraints, involving 
the hopping element $t$, etc. A 
complete analysis along these lines falls beyond
our present scope.

Another important issue concerns the physical interpretation of 
the 
Majorana fermion $\eta$, which, as one can see from (\ref{lagrangian}),
(\ref{slagrangian}), leads to  
effective 
electric-charge violating interactions on the spatial planes.  
From our two-spatial dimension point of view, such violations 
may admit the interpetation of describing 
{\it interlayer } hopping of spin and charge degrees of
freedom (hopping of real four-space-time-dimensional electrons).
In this interpretation, the gaugino $\eta$ terms in (\ref{lagrangian}), 
constitute a Majorana-spinor representation of the {\it absence}
of spin {\it and} charge at a site of the planar lattice system:
\be
   \int d\eta  e^{2i \int d^3 x {\overline \eta} \Psi ^\alpha z_\alpha + H.C.}
\label{absencepth}
\ee
The reader is advised to draw 
a comparison with the Grassmann $\chi, \chi ^\dagger $,
representation of a Wilson line (`missing spin' S ) in the 
treatment of {\it static holes} in refs. \cite{Sha,dor}:
\be
   \int d\chi ^\dagger d\chi  e^{-iS \int dt \sum _{i} (-1)^i \chi _i ^\dagger 
\chi _i a_0 (i, t) } 
\label{spin}
\ee
where $a_0$ is the temporal component 
of the gauge potential of the $CP^1$ $\sigma$-model, describing 
spin excitations in the antiferromagnet. 
From this point of view, 
the existence of $N=1$ supersymmetry in the doped 
antiferromagnets necessitates {\it interplanar couplings}, through 
hoping of spin and charge degrees of freedom (electrons) 
across the  planes. 

\section{Dynamical Mass Generation and N=1 Supersymmetry} 

Next, we proceed to discuss the dynamical scenario for 
fermion mass generation. 
First, we note that 
dynamical-mass generation (pairing) in 
non-supersymmetric models, with  combined Gross-Neveu and Thirring 
four-fermion interactions, is possible in three space-time dimensions. 
By using a four-component fermion formalism one obtains consistent 
solutions to the Schwinger-Dyson (SD) equations, with non-zero mass, 
which conserve parity and time-reversal 
invariance~\cite{mixed,mixed2}~\footnote{Note that 
theories with four-fermion
interactions are not in general vector-like, and hence the theorems 
of ref. \cite{Vafa}, for absence of spontaneous violation of parity 
and time-reversal symmetry due to energetics, cannot apply.
However, 
in our superconducting model,  
integrating out the magnon fields one 
obtains~\cite{Polyakov,ruiz} 
a dynamical gauge theory in the infrared. It is in this sense that we
are interested only in  parity-conserving mass gaps, which from the 
point of view of the (low-energy) effective gauge theory, 
are the energetically preferable configurations~\cite{Vafa}.}.  
In ref. \cite{mixed2} it was shown that 
in 
models with mixed Thirring and 
Gross-Neveu interactions, it is essentially the  Gross-Neveu coupling $g_{GN}$ 
which determines the critical behaviour (critical flavour number) 
of the theory, in a large N expansion. For $g_{GN} > g^c_{GN}$, 
where $g_{GN}^c$ is the critical coupling of the (2+1)-dimensional
Gross-Neveu model~\cite{GNmodel}, the system is dominated by the 
Gross Neveu interaction, while for $g_{GN} < g^c_{GN}$,  
the system becomes Thirring like~\footnote{In this latter 
case we should point out that 
the non-trivial ultraviolet fixed point, found in the 
numerical studies of \cite{deldebbio}, might be related -
under some sort of ultraviolet-infrared duality -
to the non-trivial 
infrared fixed point of the three-dimensional $QED$, argued in 
\cite{aitch}.}.

We now argue that qualitatively the mass -generation 
phenomenon cannot be affected 
by the 
presence of supersymmetric partners of the fermion fields. 
Indeed, the only extra terms in the lagrangian (\ref{slagrangian})
that 
could affect the dynamical mass generation are 
the terms mixing bosons and fermions,  
${\overline z}\partial _\mu z {\overline \Psi} \gamma ^\mu \Psi $. 
However, at the level of the 
effective action 
obtained from (\ref{slagrangian}) 
by path-integrating out the  
$z$ fields, the leading order contributions  
in a (low-energy) derivative expansion, are  
of order: $\int d^3x {\overline \Psi}\gamma _\mu \Psi 
[(\partial ^2  g^{\mu\nu} -\partial ^\mu \partial ^\nu)/m_Z^2 ]
{\overline \Psi} \gamma _\nu \Psi$.
Such interactions constitute irrelevant operators 
in a renormalization group sense, even at large fermion 
flavour numbers $N$, 
and hence do not affect the 
fixed-point structure of the theory, responsible for mass generation,
which is thus 
determined by the 
four-fermi terms~\footnote{We 
note that, in a large-flavour-number, $N$, treatment,   
these four-fermi operators become  
renormalizable, thereby leading to 
non-trivial ultraviolet fixed-point structures~\cite{GNmodel,mixed,mixed2}.}.

Within the context of dynamical mass generation, 
it is important to remark that in supersymmetric models  
dynamical mass generation can occur 
in a way compatible with unbroken supersymmetry only if 
the effective potential vanishes. This is a result of the equality 
of the fermion and boson masses, $m_Z=m_f=m$. 
In non-supersymmetric theories it is the minimization 
of the effective potential 
that selects the non-trivial solution of the Schwinger-Dyson (SD) analysis 
for the dynamical fermion 
mass. 
In contrast, as we shall argue below, in our supersymmetric case
it is 
the {\it quantum} effective action, and {\it not} the effective potential, 
which is responsible for such a selection. 
The situation is similar to what happens in the two-dimensional 
supersymmetric $O(3)$ $\sigma$-model~\cite{alvarez}. 
In that model,  
as a result of 
a constraint similar to (\ref{constr}),
consistency among the 
supersymmetry Ward identities, obtained from the quantum effective action, 
selects 
the non-trivial solution for the dynamical masses, obtained from a SD 
analysis~\cite{pisarski}. 
Below 
we shall not give the details, but we shall present  
the main arguments, which will be sufficient for our purposes 
in this letter.  
For simplicity we consider one ``complex'' superfield
$Z$ and work with its real components (\ref{complex}). 
The masses of the scalars $\phi _i$, $i=1,2$ and the 
Majorana spinors $\chi _i$ are related by the Supersymmetry 
Ward identity:
\be
<T\{ \chi _i (x), {\overline \chi} _j (0) \} >_o=(i\nd{\partial}
+ m)<T \{ \phi _i (x), \phi _j (0)>_o 
\label{ward}
\ee
where $ < \dots >_o $ denote correlators in the non-interacting theory. 

On the other hand, it is known that the fields:
\be
\left( -F_i = -\frac{\phi _i}{2}({\overline \chi}_j \chi _j ),~~ 
i\nd{\partial}\chi _i,~~ \partial ^2 \phi _i \right) 
\label{kinetic}
\ee
constitute real superfields $T_i$, the kinetic multiplets of $\Phi _i$.
Therefore, the vacuum expectation values of the components of the superfield: 
\be
\phi _i T_i = \left( -\phi _i F_i ,~~ \chi _i F_i-i\phi _i \nd{\partial} 
\chi _i ,~~ -\phi _i \partial ^2 \phi _i + 
i {\overline \chi }_i \nd{\partial} \chi _i + F_i F_i \right) 
\label{phiTi}
\ee
will be related by the supersymmetry Ward identities. 

Using the equations of motion and the constraint (\ref{sconstr}) 
this superfield can be written as: 
\be
  \phi _i T_i = \left( \nu,~~ \lambda,~~\alpha - \nu ^2 \right) 
\label{ftcomp}
\ee
where 
\bea
&~& \nu = -\frac{1}{2} {\overline \chi }_i \chi _i \nn \\
&~& \lambda = -i \phi _i \nd{\partial} \chi _i \nn \\
&~& \alpha = \partial _\mu \phi_i \partial ^\mu \phi_i 
\label{nal}
\eea
Then, the corresponding supersymmetry Ward identities become~\cite{alvarez}:
\bea
&~& S_{\lambda} (p) - (\nd{p}-2m) D_\nu (p^2) =0 \nn \\
&~& \nd{p} S_\lambda (p) -  D_\alpha (p^2) 
+ 2m (\nd{p} - 2m) D_\nu (p^2)  = 0
\label{wardident}
\eea
where $D_\nu, D_\alpha $ are the two-point Green's functions of $\nu $ and 
$\alpha$ fields respectively, and $S_\lambda$ is the corresponding 
spinorial Green's function of $\lambda$. Note that the equations 
of motion, obeyed by the Green's functions, have been used 
in deriving the identities above. 

In the context of the pure Gross-Neveu model in three space-time 
dimensions, one can compute the effective propagators 
by extending the two-dimensional analysis of ref. \cite{alvarez},
in a straightforward manner. For instructive purposes
we shall derive explicitly the $D_\alpha $ propagator, pertaining to 
the Lagrange multipliers $\alpha (x)$ 
implementing the constraint (\ref{sconstr}). 
Expanding about the vacuum $<\alpha (x) > =m^2$, 
$\alpha (x) = <\alpha (x)> + \alpha ' (x)$,  
and performing the $z$ integration one arrives at an 
effective action 
\be
   S_{eff,\alpha} =\int d^3x {\rm Tr}{\rm ln}[\partial ^2 + m^2 + \alpha '(x) ] 
\ee
The quadratic term in $\alpha '(x)$ 
determines the effective propagator $D_\alpha $ 
of the quantum field $\alpha '$. Passing onto a Fourier
space one obtains: 
\be
     S_{eff,\alpha}^{(2)} \sim 
\int \frac{d^3p}{(2\pi)^3}\int \frac{d^3k}{(2\pi)^3}\frac{1}{(k+p)^2 - m^2}  
{\tilde \alpha }'(-p) \frac{1}{k^2 - m^2 } {\tilde \alpha}'(p)
\label{sa}
\ee
where 
$m$ is the dynamically-generated mass for (both) scalars and 
fermions (due to supersymmetry).
From this, the propagator $D_\alpha (p)$ 
is obtained immediately.
Its $p=0$ limit is given by: 
\be
D^{-1}_\alpha (0) \sim \int d^3k \frac{1}{(k^2 - m^2)^2} \sim 1/m
\ee
\pr
In a similar manner one determines the rest of the Green's functions
appearing 
in (\ref{wardident}). For the Green's function $D_\nu$, 
associated with the linearized 
Gross-Neveu interactions, 
we note that the quantum 
corrections have been calculated in ref. \cite{GNmodel}, where a non-trivial 
ultraviolet fixed point structure has been revealed in a 
large-fermion-flavour
number, $N$, framework.

With these in mind, one obtains the 
following results for the pertinent Green's functions, to leading order in 
$1/N$ expansion:   
\bea 
&~&D_\alpha ^{-1} (0) \sim \frac{1}{m} + {\cal O}(\frac{1}{N})\nn \\
&~& D_\nu (0)^{-1} 
\sim m \left(1 + {\cal O}[\frac{1}{N}{\rm ln}(\Lambda /m)]\right) \nn \\
&~& S_\lambda ^{-1}(0) \sim \int \frac{d^3k}{(2\pi)^3} 
{\frac{1}{(k^2 - m^2)(\nd{k} -m)}} + {\cal O}(\frac{1}{N})
\sim 
{\rm non-zero~const}  + {\cal O}(\frac{1}{N}) 
\label{gnward}
\eea
where 
$\Lambda $ is an ultraviolet cut-off mass.
In the above formulas 
factors of the dimensionful coupling constant $g_1^2$ 
are understood where appropriate. Moreover, for our purposes in
this work
the 
detailed form of the ${\cal O}(\frac{1}{N}) $ corrections 
will not be 
important. 

From the Ward identities (\ref{wardident}), 
on the other hand, one has: 
\bea 
&~& 4m^2 D_\nu (0) = -
D_\alpha  (0) \nn \\
&~& 2m D_\nu (0) = - S_\lambda (0) 
\label{dnda}
\eea
Then, on account of (\ref{gnward}),  
we see that the first of the identities (\ref{dnda}) 
is satisfied identically to this order 
in 1/N, but one cannot exclude 
the 
trivial solution $m=0$. 
Such an exclusion comes from the 
second of the identities
(\ref{dnda}),  
due to the 
structure of $S_\lambda $. The so-selected 
non-trivial solution for $m$, 
must be the one 
satisfying the SD equations~\cite{GNmodel},
by consistency. A non-trivial verification
of this will come by including the subleading 1/N corrections.   
The reader should keep note of 
the crucial r\^ole of the $CP^1$ constraint (\ref{sconstr})
in the above selection of the non-trivial SD mass gap by the quantum 
effective action
of the $N=1$-supersymmetric model~\cite{alvarez,pisarski}. 

In the context of our model, involving both Gross-Neveu and
Thirring interactions, a similar analysis goes through,
with complexities coming from the non-linear realization of 
supersymmetry, and the new interactions in (\ref{slagrangian}).
Such deviations
from the pure Gross-Neveu case, however, are in favour 
of the necessity of a non-zero mass gap, in order to 
fulfill the supersymmetry Ward identities. 
A detailed analysis along the above lines 
will be presented in a forthcoming publication.
For the purposes of this note we restrict oursleves only to 
pointing out some subtleties,  
associated with the anomalous breaking of the fermion number 
in our model (c.f. figure 1). 
Indeed, after the $z$ integration, 
and the implementation of the constraint as above, there 
are extra terms coupling fermions and $\alpha $ multiplier fields
in the effective action. One of them involves the divergence of the 
fermion current (after appropriate partial integrations in the action):
\be 
    S_{eff}^{\chi,\alpha}~ \ni~ -\int d^3x 
{\rm Tr}\frac{\partial _\mu ({\overline \chi }_i \gamma ^\mu 
\chi_i)}{(\partial ^2 + m^2)^2}\alpha (x)
\label{psialpha}
\ee
If the fermion current ${\overline \chi} _i \gamma ^\mu 
\chi_i$ was conserved, then the 
gauge Ward identity would imply decoupling of this term from the physical 
correlators. However, as we mentioned above, there are anomalies in the model, 
in the massive phase for the fermions, associated with 
one-loop graphs of figure 1~\cite{RK,dor,fm}. 
Such anomalous terms should be 
properly taken into account in a detailed analysis
of dynamical mass generation in our supersymmetric model, 
but we do not expect them to affect 
the selection of the non-trivial solution 
of the SD equations,
characterising the pure Gross-Neveu case, studied in detail above. 

An important additional comment 
concerns the kind of the three-dimensional 
dynamically-generated  mass.
At present, this seems to depend crucially 
on the relative sign of the mass, between the fermion species. 
In our analysis above, we have used 
a single superfield $Z$, whilst in 
our $SU(2)$ model there are two such superfields. 
The SD analysis 
can be extended in that case straightforwardly, but  
alone it cannot make 
a selection among the two possible signs of the mass
for these two superfields.
Since the four fermion theories are not vector like,  
one does not have at first sight a way of energetically selecting the 
parity-conserving mass configuration. 
However, as we mentioned previously, the 
fact that the low-energy integration of magnon fields 
makes the model equivalent to a (vector-like) 
gauge theory with fermionic matter, is suggestive 
of the exclusion of 
the parity violating mass, 
on the basis of the theorems of ref. \cite{Vafa}.

\section{Instanton Effects, Supersymmetry and Superconductivity} 

A final issue we would like to address
concerns the 
{\it exactness of
superconductivity} in the presence of {\it non perturbative } effects. 
In the context of the $SU(2) \times U_S(1)$ theory~\cite{fm}, 
superconductivity is associated with the 
masslessness of the $B_\mu ^3$ gauge boson of the 
unbroken $U(1)$ subgroup of the $SU(2)$ 
group, in the massive fermion phase~\cite{fm,dor}. We now remark that, 
due to the compactness of the pertinent gauge group, instanton 
configurations - which, in 2+1 dimensions, are like monopoles - 
may give the $U(1)$ gauge boson 
a small mass. In the dilute-instanton-gas approximation,  
in non-supersymmetric theories, this mass is of order~\cite{polyakovinst}:
\be
     m_{B_3} \simeq e^{-S_0} 
\label{oneinst}
\ee
where $S_0$ is the one instanton action. 
Such a small mass would destroy the exactness of the model's 
superconductivity, as we remarked earlier.

We shall argue in this section that supersymmetry favours 
superconductivity, by further
suppressing the instanton contributions to the $B_\mu^3$ gauge boson mass,
as compared to the non-supersymmetric case. 
To this end, we first recall that 
a dynamical gauge theory is obtained in our model by 
integrating out $z$ and $\Psi $ fields~\cite{Polyakov}.
In a non-supersymmetric theory, upon coupling to 
external electromagnetism, such a procedure leads, 
in the massive fermion phase, to the 
standard London action for superconductivity~\cite{dor}. 
In our case, this procedure leads to a supersymmetric 
gauge theory $U(1) \times U_S(1)$. 
Indeed, by integrating out $z$ fields one obtains Maxwell kinetic terms 
for the gauge fields, in the phase where 
the magnon fields 
are massive~\cite{Polyakov}. In our supersymmetric theory, the Yukawa coupling 
of the gaugino $\eta$ to $\Psi $ and $z$ 
fields results, after the $z$,$\Psi$ 
integration in Majorana kinetic terms for $\eta$, as required by $N=1$ 
supersymmetry. 
This  can be readily seen by a one loop computation, in analogy with 
the bosonic $z$ part~\cite{Polyakov}. The relevant graphs, in the 
massive phase $m_Z=m_f=m$ (due to supersymmetry),  
result in the following integral:

\be 
\gamma ^\mu \int d^3 k \frac{k_\mu + p_\mu }{[(k+p)^2 - m^2][ k^2 - m^2]} 
\sim \frac{\nd{p}}{2 m} 
\label{integral}
\ee
yielding a Majorana kinetic term 
$\frac{i}{2m}{\overline \eta}\nd{\partial} \eta$  
for the gaugino. One can easily verify the 
manifest $N=1$ supersymmetry between this term and the corresponding 
Maxwell terms $-\frac{1}{4m} F_{\mu\nu}^2 $, obtained by the 
$z$ and $\Psi $ integration~\cite{Polyakov,dor}. 

We now 
remark 
that 
in three dimensional 
supersymmetric gauge theories 
it is known~\cite{harvey} that  
supersymmetry cannot be broken, due to the fact that the Witten 
index $(-1)^F$, where $F$ is the fermion number, is always non zero. 
Thus, in supersymmetric theories the presence of instantons 
should give a small mass, if at all, in both the gauge boson and the 
associated gaugino. Although at present there is no rigorous proof 
of this fact, however, the arguments of ref. \cite{harvey}
indicated that the resulting masses will be even more suppressed 
than the corresponding ones in the non supersymmetric case, 
\be
    m_{B_3}=m_{\eta} = e^{-2S_0} 
\label{sinstmass}
\ee
with $S_0$ the one-instanton action. 

We should point out,
however, that 
there is an alternative scenario~\cite{harvey},
which could be in operation in our superconducting model. 
It is possible that 
supesymmetry is broken by having the system in a `false' vacuum,
where the gauge boson remains massless, even in the presence of 
non perturbative configurations, while the gaugino acquires a 
small mass, through non perturbative effects. 
The life time, however, of this false vacuum is very long~\cite{harvey}, 
and hence superconductivity can occur, in the sense that 
the system will remain in that false vacuum 
for a very long period of time, longer than any other time scale 
in the problem.

Whichever of the two scenaria is realized in the model, 
from a condesed-matter point of view 
the important conclusion, obtained  from the above analysis,  
is 
that the coupling of the superconducting planes
due to interlayer electron hopping, associated with the presence of the 
gaugino field $\eta$,  
helps stabilizing 
superconductivity, which otherwise would be jeopardized by 
non-perturbative effects.

\section{Discussion} 

In this work we have demonstrated the
possibility of N=1 Supersymmetric gauge theories
in the context of a spin-charge separation ansatz
of the $SU(2) \times U_S(1)$ gauge model of \cite{fm}. 
Such models may 
be relevant for the physics of
superconducting gaps which open up 
at the nodes of a d-wave gap
of high-$T_c$ cuprates. The supersymmetry was achieved 
without introducing unphysical degrees of freedom. 
However 
it necesitates the coupling of the superconducting planes.
Its presence, seems to suppress the effects of instantons 
of the gauge field $SU(2)$, which could jeopardise 
superconductivity in the model.  
As far as the lattice system is concerned the  supersymmetry 
is achieved {\it modulo} 
irrelevant operators in a renormalization-group sense. 
This may imply that 
our considerations in this work 
might also be relevant to the construction of more general 
supersymmetric gauge 
theories on the lattice, 
in the sense of obtaining supersymmetric continuum theories 
by droping possibly non-supersymmetric, renormalization-group  
irrelevant operators.

We believe that 
our work may prove useful towards 
an exact discussion 
of phase diagrams of three-dimensional  
effective gauge models of antiferromagnetic 
superconductors, 
via the analysis of 
the quantum 
moduli space of gauge theories, in the spirit 
of Seiberg and Witten~\cite{seiberg}. 
In this respect, we note that 
$N=1$ supersymmetry in three dimensions, which 
we have considered here as the minimal way of supersymmetrization of a doped 
spin-charge separated antiferromagnet, without the introduction of extra degrees
of freedom, cannot yield exact results. It is the $N=4$ 
supersymmetry in three-dimensional gauge theories which
can produce such results. In three dimensions, 
$N=2$ theories may also allow  for some exact 
results, in connection with the geometry 
of their quantum moduli space~\cite{intrilligator}.  

At present, our physical 
understanding for a condensed-matter spin-charge 
separated model exhibiting $N=2$ supersymmetry 
is not complete. One might speculate that, since $N=2$
three-dimensional supersymmetric theories 
are obtained~\cite{intrilligator} 
by dimensional reduction of 
four-dimensional $N=1$ supersymmetric theories,  such models
might have some relevance to a possible 
extension of the 
ideas in our work
beyond the planar structures. 
We shall present a more detailed study of 
such models in a future work~\cite{lahanas}. We should remark however 
that, 
as far as the spinon and holon degrees of freedom are concerned, the 
extension 
to $N=2$ supersymmetry is immediate, with no extra doubling of degrees
of freedom.  The novel feature, compared to the N=1 case, is 
the presence of 
a {\it Dirac-like} gaugino. Due to its Dirac nature, the gaugino
may now carry non-trivial charge under the external electromagnetism, 
and thus the effective action conserves the 
electric charge, in contrast to the present situation 
with a Majorana 
gaugino. In view of the aforementioned embedding of 3-dimensional N=2
theories in 4-dimensional  N=1 supersymmetric theories,  
this possibility of 
conservation of the electric charge 
may be related 
to the exact conservation of electric charge in four dimensional 
space times. From the point of view of dynamical mass generation,
we should remark that, at first sight, the $N=2$ 3-dimensional models 
appear not to generate a dynamical mass. This is due to the fact that such 
theories are obtained from $N=1$ 4-dimensional models, where 
claims have been made~\cite{clark} that
non-renormalization theorems in the supersymmetric SD equations 
yield only the trivial solution for the mass.
However, such claims have been questioned recently~\cite{apelrec}. 
From our point of view 
we consider the issue as still open.  

Another comment we would like to make concerns the fate of 
supersymmetry at finite temperatures. 
We expect the supersymmetry 
to 
be broken at finite 
temperatures, which results in different masses for spinon and holons,
a situation probably met in realistic cases. However, 
even in such a case of broken supersymmetry, 
the existence of a supersymmetric vacuum at zero temperatures 
is useful in providing some exact information 
about the phase 
diagram along the lines mentioned above. 

Before closing we should also stress 
that our results apply even to 
one-dimensional chains of holons, which may characterize 
certain underdoped cuprates in the so-called stripe phase~\cite{stripes}.
Such systems appear to be described by a spin-charge separated 
phase, where 
the holon degrees of freedom lie 
on one-space dimensional stripes (chains), 
spatially
separated by regions of zero doping. 
As discussed in ref. \cite{Sha}, spin charge separation 
in one (spatial) dimensional antiferromagnetic 
models leads to gauge theories of Dirac fermions coupled
to a $CP^1$ $\sigma$-model.
The continuum action is similar in form, but in two space-time
dimensions, with the 
action (\ref{su2action}). 
In such a case, the resulting $N=1$ supersymmetric  extension will 
again involve combined Gross-Neveu and 
Thirring interactions. Such (1+1)-dimensional models have been 
studied previously in the literature~\cite{alvarez}.
As far as supersymmetry and dynamical mass 
generation are concerned, 
such models 
share the same
qualitative features
as their (2 + 1)-dimensional counterparts,
discussed here. 
The gauginos in such one-dimensional theories  
could then describe (effective)  
electron hopping across the chains. At present we are agnostic
as to 
whether such supersymmetric spin-charge separated 
models play any crucial r\^ole on the physics 
of the stripe phase of the 
underdoped cuprates.

\section*{Acknowledgements} 

G.A.D. and B.C.G. wish to thank the Department of (Theoretical) Physics 
of Oxford University for the hospitality.

\end{document}